\documentclass[runningheads]{llncs}

\usepackage[utf8x]{inputenc}
\usepackage{graphicx}
\usepackage{multirow}
\usepackage{subfigure}
\usepackage{color}
\usepackage{enumitem}

\newcommand{\csa}{\mathsf{CSA}}
\newcommand{\icsa}{\mathsf{iCSA}}
\newcommand{\wm}{\mathsf{WM}}

\newcommand{\tcsa}{\mathsf{CTR}}    %%UNIFICAR AL FINAL
\newcommand{\repres}{\mathsf{CTR}}   %%UNIFICAR AL FINAL

\begin{document}
\title{Compact Trip Representation over Networks
  %
  %\title{Compact Spatio-Temporal Representation for Trips over Networks
   \thanks{
     \footnotesize{
     Funded in part by European Union's Horizon 2020 research and innovation programme
     under the Marie Sklodowska-Curie grant agreement No 690941 (project BIRDS).
     N. Brisaboa, A. Fari\~na, and  D. Galaktionov are funded  by  MINECO (PGE, CDTI, and FEDER)
  [TIN2013-46238-C4-3-R, TIN2013-47090-C3-3-P, TIN2015-69951-R, IDI-20141259, ITC-20151305, ITC-20151247];
     by ICT COST Action IC1302; and by Xunta de Galicia (co-funded with FEDER) [GRC2013/053].
     A. Rodr\'\i guez is funded by Fondecyt 1140428 and the Complex Engineering Systems Institute (CONICYT: FBO 16)
     %G. Navarro is supported by Millennium Nucleus Information and Coordination in Networks ICM/FIC P10-024F, Chile.
     }
   }
}

\titlerunning{Compact Trip Representation over Networks}

\author{Nieves R. Brisaboa\inst{1} \and
    Antonio Fari\~na\inst{1} \and
    Daniil Galaktionov\inst{1} \and\\
%   Gonzalo Navarro \inst{2} \and
    M. Andrea Rodr\'iguez\inst{2} %\\
    %\url{\{brisaboa,fari,d.galaktionov\}@udc.es} - \url{andrea@udec.cl}
    }

\authorrunning{Brisaboa et al.}
\authorrunning{N. Brisaboa, A. Fari\~na, D. Galaktionov, and A. Rodr\'iguez}

\institute{Database Laboratory, University of A Coru\~na, Spain \and % University of Chile \and
    Department of Computer Science, University of Concepci\'on, Chile}

\maketitle

%%%%%%%%%%%%%%%%%%%%%%%%%%%%%%%%%%%%%%%%%%%%%%%%%%%%%%%%%%%%%%%%%%%%%%%%%%%%%%%%%%%%%%%

%%%%%%%%%%%%%%%%%%%%%%%%%%%%%%%%%%%%%%%%%%%%%%%%%%%%%%%%%%%%%%%%%%%%%%
%%%%%%%%%%%%%%%%%%%%%%%%%%%%%%%%%%%%%%%%%%%%%%%%%%%%%%%%%%%%%%%%%%%%%%
\begin{abstract}

We present a new {\em Compact Trip Representation} ($\repres$) that
allows us to manage users' trips (moving objects) over networks.
These could be public transportation networks (buses, subway,
trains, and so on) where nodes are stations or stops, or road
networks where nodes are intersections. $\repres$ represents the
sequences of nodes and time instants in users' trips. The spatial
component is handled with a data structure based on the well-known
Compressed Suffix Array ($\csa$), which provides both a compact
representation and interesting indexing capabilities. We also
represent the temporal component of the trips, that is, the time
instants when users visit nodes in their trips. We create a sequence with these
time instants, which are then self-indexed with a balanced Wavelet Matrix ($\wm$). This
gives us the ability to solve range-interval queries efficiently. We
show how $\repres$\ can solve relevant spatial and spatio-temporal
queries over large sets of trajectories. Finally, we also provide
experimental results to show the space requirements and query
efficiency of $\repres$.

\end{abstract}

%%%%%%%%%%%%%%%%%%%%%%%%%%%%%%%%%%%%%%%%%%%%%%%%%%%%%%%%%%%%%%%%%%%%%%
\section{Introduction}
%%%%%%%%%%%%%%%%%%%%%%%%%%%%%%%%%%%%%%%%%%%%%%%%%%%%%%%%%%%%%%%%%%%%%%
Current technology allows us to capture data about the usage of
transportation networks whose analysis could have an important
impact on improving the quality of services. Data about the
origin and destination of passengers of train  services can be
directly captured when selling tickets. Using more sophisticated
technology, the movement of people or vehicles over networks of
streets or roads can be collected from the mobile phone signals.
Even more, nowadays many cities (from London to Santiago of Chile)
provide smartcards to the users of their public transportation
network. These smartcards (that can be recharged with money) allow
users to pay the entrance to subways and buses. Even though there
typically exists only a card reader in the entrance to the network
(i.e., there is no control at exits or in middle stops), it is
possible to know how people actually use the public transportation
by collecting the entrance and estimating the destination (e.g., 
as the entry point for the return trip) and the traversed
stops (the shorter path among stops used as the entrance and
exit)~\cite{Munizaga20129}. In all scenarios, the massive data about
trips makes the problem of storing and efficiently accessing  data
about trips a challenging computational problem.

This paper presents a compact and self-indexed data structure to
represent trips over networks,  which could be public transportation
networks where nodes are stations or stops, or road networks where
nodes are intersection points.

Although there exist proposals of data structures for moving
objects, they have addressed typical spatio-temporal queries such as
time slice or time interval queries that retrieve trajectories or
objects that were in a spatial region at a time instant or during a
time interval. They were not designed to answer queries that are
based on counting occurrences such as the number of trips starting
or ending at some time instant in specific stops (nodes) or the
top-k most used stops of a network during a given time interval,
which are more meaningful queries for public-transportation or
traffic administrators. Our proposal ($\repres$) is oriented to efficiently
answering these types of queries, and it differs from previous approaches
in the use of compact self-indexed data structures to represent
the big amount of trips in compact space.
%It is important to emphasize that our goal is to provide an index of the trips
%for batch processing, without inserting new trips.
 It is important to emphasize that our goal is to provide an indexed 
 representation for a static collection of trips in order to allow
 an efficient batch processing of such data.

 $\repres$ combines two well-known data structures. The first one,
initially designed for the representation of strings, is 
Sadakane's Compressed Suffix Array ($\csa$) \cite{Sad03}. The second
one is the Wavelet Matrix ($\wm$) \cite{CNO15}. To make the
use of the $\csa$ possible in this domain, we define a trip or trajectory of a moving object
over a network as the temporally-ordered sequence of the nodes the trip
traverses.  An integer $id$ is assigned to each node such that a trip
is a string of nodes' $id$s. Then a $\csa$, over the concatenation of
these strings (trips) is built with some adaptations for this
context. In addition, we discretize the time in periods of fixed
duration (i.e. timeline split into 5-minute instants) and each time
segment is identified by an integer $id$. In this way, it is possible
to store the times when trips reach each node by associating the
corresponding time $id$ with each node in each trip. The sequence of
times for all the nodes within a trip is self-indexed with a $\wm$
to efficiently answer spatio-temporal queries.

We experimentally tested our proposal using two sets of synthetic
data representing trips over two different real public
transportation systems. Our results are promising because the
representation uses only around  30\% of its original size and
answers spatial  and spatio-temporal queries in microseconds. No
experimental comparisons with classical spatial or spatio-temporal
index structures are possible, because none of them were designed to
answer the types of queries in this work. Our approach can  be
considered as a proof of concept that opens new application
domains for the use of $\csa$ and $\wm$, creating a new strategy for
exploiting trajectories represented in a self-indexed way.

The organization of this paper is as follows. Section~\ref{sec:prevwork} reviews previous works on
trip representations. It also makes reference to the
$\csa$ and $\wm$, upon which we develop our proposal.
 Section~\ref{sec:transnet_repr} shows how  $\repres$ represents the spatial component  and Section~\ref{sec:time_repr} the  temporal component of trips.
Section~\ref{sec:tcsaqueries} presents the  relevant queries that are solved by $\repres$ and
Section~\ref{sec:experiments} gives our experimental results.
%gives the experimental evaluation of $\repres$., showing  its efficiency in space and time.
Finally, conclusions and future work are discussed in Section~\ref{sec:conclusions}.

%%%%%%%%%%%%%%%%%%%%%%%%%%%%%%%%%%%%%%%%%%%%%%%%%%%%%%%%%%%%%%%%%%%%%%%%%%%%%%%%%%

%%%%%%%%%%%%%%%%%%%%%%%%%%%%%%%%%%%%%%%%%%%%%%%%%%%%%%%%%%%%%%%%%%%%%%
\section{Previous Work} \label{sec:prevwork}
%%%%%%%%%%%%%%%%%%%%%%%%%%%%%%%%%%%%%%%%%%%%%%%%%%%%%%%%%%%%%%%%%%%%%%

%%%%%%%%%
%\subsection{Previous work in the  representation of trajectories}
\noindent
\textbf{Trajectory indexing.} Many data structures have been proposed to support efficient query capabilities on collections of
trajectories. We refer to~\cite[Chapter 4]{DBLP:books/sp/PelekisT14} for a comprehensive and
up-to-date survey on data management techniques for trajectories of moving objects. We can broadly
classify these data structures into two groups: those that index trajectories in free space and
those that index trajectories %that are 
constrained to a network. The 3D R-tree (an extension of the
classical R-tree spatial index~\cite{DBLP:conf/sigmod/Guttman84}), the
TB-tree~\cite{DBLP:conf/vldb/PfoserJT00}, and MV3R-tree~\cite{DBLP:conf/vldb/PapadiasT01} are
examples of the former, whereas the FNR-tree~\cite{DBLP:conf/ssd/Frentzos03}, the
MON-tree~\cite{DBLP:journals/geoinformatica/AlmeidaG05}, and
PARINET~\cite{DBLP:journals/vldb/PopaZOBV11} are examples of the latter.
While the former type of structures could also apply over networks, the second type exploits the
constraints imposed by the topology of the network to optimize the data structure.
From them, PARINET is the most efficient alternative~\cite{DBLP:journals/vldb/PopaZOBV11}. %
%
% It divides trajectories into segments associated to the
% underlying roads in a network, and for each segment keeps: 
% the trajectory-id, the road-id for the segment, and the position and times of the starting and ending points. Both roads and
% trajectory segments are stored in a data base.
% At query time, the segments of the trajectories can be easily grouped by road, and a temporal 
% B$⁺$-tree %(relying on the DBMS) 
% index (there is one index for each road)
% is used  to filter out candidate segments matching time constraints.
%
%
It partitions trajectories into segments from an underlaying road network,
and then adds one temporal B$⁺$-tree to index the trajectory segments from each road. 
Those indexes permit us to
filter out candidate trajectory segments matching time constraints at query time.

All previous data structures were designed to answer spatio-temporal queries, where the space and time are the main
filtering criteria. Examples of such queries are: {\em retrieve trajectories that crossed a region
within a time interval}, {\em retrieve trajectories that intersect}, or {\em retrieve the $k$-best connected
% * <galaktionov@gmail.com> 2016-05-19T09:09:47.441Z:
%
% > All previous data structures were designed to answer spatio-temporal queries, where the space and time are the main
% > filtering criteria. Examples of such queries are: {\em retrieve trajectories that crossed a region
% > within a time interval}, {\
%
% ah, y puedes anadir comentarios...
%
% ^.
trajectories} (i.e., the most similar trajectories in terms of a distance function).
Yet, they could not easily support queries such as {\em number of trips starting in X and ending at Y}.

%Although existing data structures for trajectories support efficient
%query processing on  large datasets, they can have  limitations to
%deal with current massive data.
 %In this context, data compression techniques has been explored in the past to help in handling  the  problem of massive data.
The application of data compression techniques has
been explored in  the context of massive data about trajectories. The work by Meratnia and de
By~\cite{DBLP:conf/edbt/MeratniaB04} adapts a classical
simplification algorithm by Douglas and Peucker to reduce the number
of points in a curve and, in consequence, the space use to represent
trajectories. Ptomaias \textit{et
al.}~\cite{DBLP:conf/ssdbm/PotamiasPS06} use concepts, such as speed
and orientation, to improve compression. Both techniques work for
trajectories in free space.
%When we know that object movements are constrained to a
%network, we can do even better. This aspect has been
%explored
%in~\cite{DBLP:journals/josis/RichterSL12,DBLP:journals/jss/KellarisPT13,DBLP:conf/w2gis/FunkeSSS15}.
%Those works focus mainly in how to represent trajectories and in how to gather the location of one or more
%given  moving objects from those trajectories. Yet, they would poorly support more complex aggregated queries.
%\marginpar{porfa daniil, mira se o anterior é correcto}
%

In~\cite{DBLP:journals/josis/RichterSL12,DBLP:journals/jss/KellarisPT13,DBLP:conf/w2gis/FunkeSSS15},
they focus mainly on how to represent trajectories constrained to
networks, and in how to gather the location of one or more given
moving objects from those trajectories. %Yet, these works are also
%out of our scope as they would poorly support queries related to an
%unbounded set of objects, or more complex aggregated queries 
%oriented to exploit the data about the network usage.
%
Yet, these works are also out of our scope as they would poorly support 
queries oriented to exploit the data about the network usage such as 
those oriented to aggregate the number of trips  with a specific  spatio-temporal
pattern (e.g. Count the trips starting at stop $X$ and ending at stop $Y$ in 
working days between 7:00 and 9:00).

%In \cite{krogh2014path} an index is presented to solve a particular and interesting kind of queries called the Strict Path Queries. However, it is a database implementation and as such it brings a prohibitive waste of space for a succinct data structure like the $\repres$.

In \cite{krogh2014path}, authors use a representation of trajectories where for each edge in a trajectory both the starting and ending times are kept, and present an index called {\em NETTRA}. They used a relational database where those data are stored in a table and indexes are created in order to support a particular type of queries called {\em Strict Path Queries}. Although our $\repres$ could also deal with those types of queries, this database-oriented representation is out of our scope as they do not consider space constraints (they do not compress data nor do they consider the size of the indexes used).

%Data structures envisioned in our work must support the
%representation of trajectories on a network. Indeed, we are not
%interesting in codifying trajectories as a sequence of locations on
%a free space, but as a sequence of nodes visited on a transportation
%network.
%Also, the recent work
%in~\cite{DBLP:conf/w2gis/FunkeSSS15} encodes  a
%trajectory as a concatenated string  and  queries match
%patterns within a large concatenation of trajectories.

%Our proposal also represents a trajectory (trip) as a  text sequence.  However, unlike previous works, we designed a
%compact data structure to make
%it successful not only in time but also in space needs.
%\marginpar{Quizas aqui no haria falta meter ``Our proposal...'' del parrafo anterior}

\vspace{0.2 cm}
\noindent

\textbf{Underlying Compact Structures of $\repres$.}  Our proposal
is based on  two well-known compact structures: a Compressed Suffix
Array ($\csa$)~\cite{Sad03} and a Wavelet Matrix
($\wm$)~\cite{CNO15}. We used the variant of $\csa$ from
\cite{FBNCPR12}, where authors adapted $\csa$ to deal with large
(integer-based) alphabets and created the {\em integer-based CSA}
($\icsa$).
 %  LA SIGUIENTE FRASE ESTABA COMENTADA PERo SI ES VERAD ME GUSTRARIA
%  QUE ESTUVIERA
They also showed that the best
 compression of $\Psi$ was obtained when combining differential
 encoding of runs with Huffman and run-length encoding.

$\wm$ is a data structure originated from the Wavelet
Tree~\cite{WT03}, but requires less space and permits to make an
efficient occurrence count of a continuous range of
values~\cite{CNO15} (see Section \ref{sec:time_repr} for details).
$\wm$ provides,  as the Wavelet Tree a self-indexed representation
of symbols based on the rearrangement of their bits in different bit
maps at different levels. $\wm$ allows us to perform efficient
operations over the sequence, among other operations: $access(i)$
returns the symbol at the position $i$, $rank_{\alpha}(i)$ counts
the number of occurrences of a symbol $\alpha$ up to position $i$;
and $select_{\alpha}(j)$ gives the position of the $j$-th $\alpha$.
Those operations are implemented using the classical bit operations
$rank$ and $select$ on the underlying bitmaps and they need
$O(\log\sigma )$ time, being $\sigma$ the number of encoded symbols.

\section{Compact Trip Representation ($\repres$)}

Trips on networks are temporally-ordered sequences of nodes (referred to as the spatial component)
tagged with timestamps (referred to as the
temporal component).  We show how the proposed  {\em Compact Trip Representation} ($\repres$) combines a
Compact Suffix Array ($\csa$)  to represent the spatial component and
a Wavelet Matrix ($\wm$) to represent the temporal one.
%
%\marginpar{quizas esto se puede ``mergear''con la Sección previa {\em Underlying Compact Structures of CTR}.}

\subsection{Representing the spatial component of $\repres$ with a $\csa$}
\label{sec:transnet_repr}
%%%%%%%%%%%%%%%%%%%%%%%%%%%%%%%%%%%%%%%%%%%%%%%%%%%%%%%%%%%%%%%%

In  $\repres$, integer IDs identify  stops of the network.  The
first step to build the $\csa$ is to sort the trips. They are sorted by the first stop, then by the last stop, then by the start time of the trip, and
finally by the second, third, and successive stops.  For example, we have a dataset $\mathcal{T}$ with the following set of trips: $\{$%
$\langle 2,3,10,6 \rangle$,
$\langle 2,3,10,4,7 \rangle$,
%$\langle 2,3, \underline{10} ,4 \rangle$,
$\langle 1,2,3\rangle$
$\langle 3,10 ,5 \rangle$,
$\langle 1,2,3\rangle$
$\langle 9,8,7 \rangle$$\}$.  Let us assume that these trips start at time instants 10, 2, 0, 9, 5, 12, respectively.
Following %a simple
lexicographic order, the trip $\langle 2,3,10,4,7 \rangle$ should be before the trip $\langle 2,3,10,6 \rangle$. However, because
after the first stop, we consider the  last stop,
%before considering the second and following stops,
the trip $\langle 2,3,10,6 \rangle$ goes before
the trip $\langle 2,3,10,4,7 \rangle$. In addition, the two trips  $\langle 1,2,3\rangle$ are sorted by their starting time instants (0 and 5
respectively). %, and the one that
%starts at time instant 25 will precede the one at instant $38$.
 This sorting of the trips will allow us to answer a useful query very efficiently
(i.e., trips starting at $X$ and ending at $Y$).

%\marginpar{{OJO:\tiny Habra que cambiar Fig1 para que I e Icodes sean como en figura 2, eso obliga tambien a cambiar los valores 25 y 38 de este
%ejemplo y la secuencia de starting time intervals: 58, 168, 25, 145, 38, 96 }}

\begin{figure}[h!]
  \vspace{-0.4cm}
  \begin{center}
  {\includegraphics[width=1.00\textwidth]{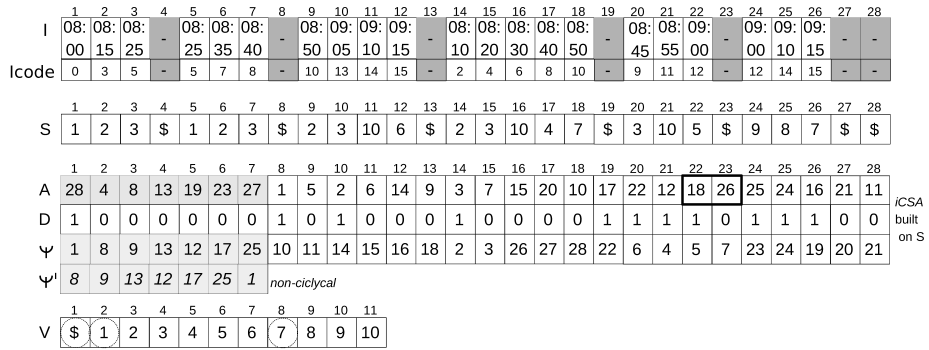}}
  \end{center}
  \vspace{-0.4cm}
  \caption{Structures involved in the creation of a $\tcsa$.}
  \label{fig:tcsa}
  \vspace{-0.5cm}
\end{figure}

%With the concatenation of ordered trips, we
We concatenate the sorted trips and construct an array $S$ where  trips are separated with a symbol \$. We also add an additional ending \$. %at
%the end   $S$.
Figure~\ref{fig:tcsa} shows the array $S$ for the running example.
Despite the standard suffix array construction in the  $\csa$ that compares two suffixes by their lexicographical order until the end of $S$,
we introduced a modification so that two suffixes are now compared considering their trips as a cycle.

% Over sequence $S$, we construct a suffix array  with a
% modification over the standard $\csa$. Instead of comparing two suffixes by their lexicographical order until the end of $S$,  two suffixes are now
% compared considering their own trips as a cycle.

Figure~\ref{fig:tcsa} depicts the structures $\Psi$ and $D$ used by the
$\tcsa$ over the trips in the dataset $\mathcal{T}$. There is also the vocabulary $V$
containing all the stops in their lexicographic order, as well as
the $\$$ symbol.  We include the sequence $S$, the
suffix array $A$, and  $\Psi$' only for clarity (they are not needed in the $\tcsa$).
 $\Psi'$  contains
the first entries of $\Psi$ from a regular $\csa$, just to explain the difference of how we build $\Psi$. For example, $A[8]=1$ points to the first stop of the first trip $S[1]$.
$\Psi[8]=10$ and $A[10]=2$ points to the second stop.  $\Psi[10]=14$ and $A[14]=3$ points to the third stop.
$\Psi[14]=2$ and $A[2]=4$ points to the ending $\$$ of the first trip. Therefore, in the standard $\csa$, $\Psi'[2]=9$ and $A[9]=5$  points
to the first stop of the second trip. However, in  $\repres$, $\Psi[2]=8$ and $A[8]=1$ points
to the first stop of the first trip. Thus, subsequent applications of $\Psi$ will allow us to cyclically
traverse the stops of the trip.
Finally, note that aligned with
sequence $S$, we could keep the times associated with the stops in each
trip with the structures $I$ and $Icode$, which are explained in the following subsection.

The definition of {\em a suffix} proposed above explains why $A[22] = 18$  is placed before $A[23]= 26$. Note that
the suffix starting at $S[18]$ is ``$7 \cdot \$ \cdot 2 \cdot 3 \dots$'' and that suffix at
$S[26]$ is ``$7 \cdot \$ \cdot 9 \cdot \dots$''. Therefore, it holds that $A[22] \prec A[23]$. However, considering
the traditional definition of a {\em suffix}, these suffixes would be ``$7 \cdot \$ \cdot 3 \cdots $''
and ``$7 \cdot \$ \cdot \$ \cdots $'' respectively,  and  $A[22] \prec A[23]$ would not hold.

%In  Figure~\ref{fig:tcsa}, $\Psi'$  contains
%the first entries of $\Psi$ from a regular \csa. For example, $A[8]=1$ points to the first stop of the first trip $S[1]$.
%$\Psi[8]=10$ and $A[10]=2$ point to the second stop.  $\Psi[10]=14$ and $A[14]=3$ point to the third stop.
%$\Psi[14]=2$ and $A[2]=4$ point to the ending $\$$ of the first trip. Therefore, $\Psi'[2]=9$ and $A[9]=5$ point
%to the first stop of the second trip. However, we managed to modify $\Psi$ so that $\Psi[2]=8$ and $A[8]=1$ point
%to the first stop of the first trip. Thus, subsequent applications of $\Psi$ will allow us to cyclically
%traverse the stops of each trip. This will be interesting at query time.

Note also that, in the shaded range $\Psi[1,7]$, the first entry is
related to terminator $\$$, whereas the next six entries  correspond
to the $\$$ symbols that mark the end of each trip in $S$ (sorted
by the starting stop, then by the ending stop, then by their initial time, and finally by the second, third and following up stops).
This property makes it very simple to find
starting %and ending
stops. For example, the ending $\$$ of the
$4^{th}$ trip is at the $5^{th}$ position (because the first $\$$
corresponds to the final $\$$ at $S[28]$). Therefore, its starting
stop can be obtained by  $\Psi[5] = 12$ and $rank_1(D,12)= 3$;
that is, the starting stop is the $3^{th}$ entry in the vocabulary.
The next stop of that trip would be obtained by $\Psi[12] = 16$
and $rank_1(D,16)=4$, and so on.

We  expect to obtain good compressibility  in $\tcsa$
due to the structure of the network, and the fact that trips that start in a given stop or simply
those going through that stop will probably share the same sequence of ``next'' stops. This will
lead us to obtain many {\em runs} in $\Psi$~\cite{NM07}, and consequently, good compression.

\subsection{Representing the temporal component of $\repres$  with a $\wm$}
\label{sec:time_repr}

\begin{figure}[h!]
  \vspace{-0.4cm}
  \begin{center}
  {\includegraphics[width=1.0\textwidth]{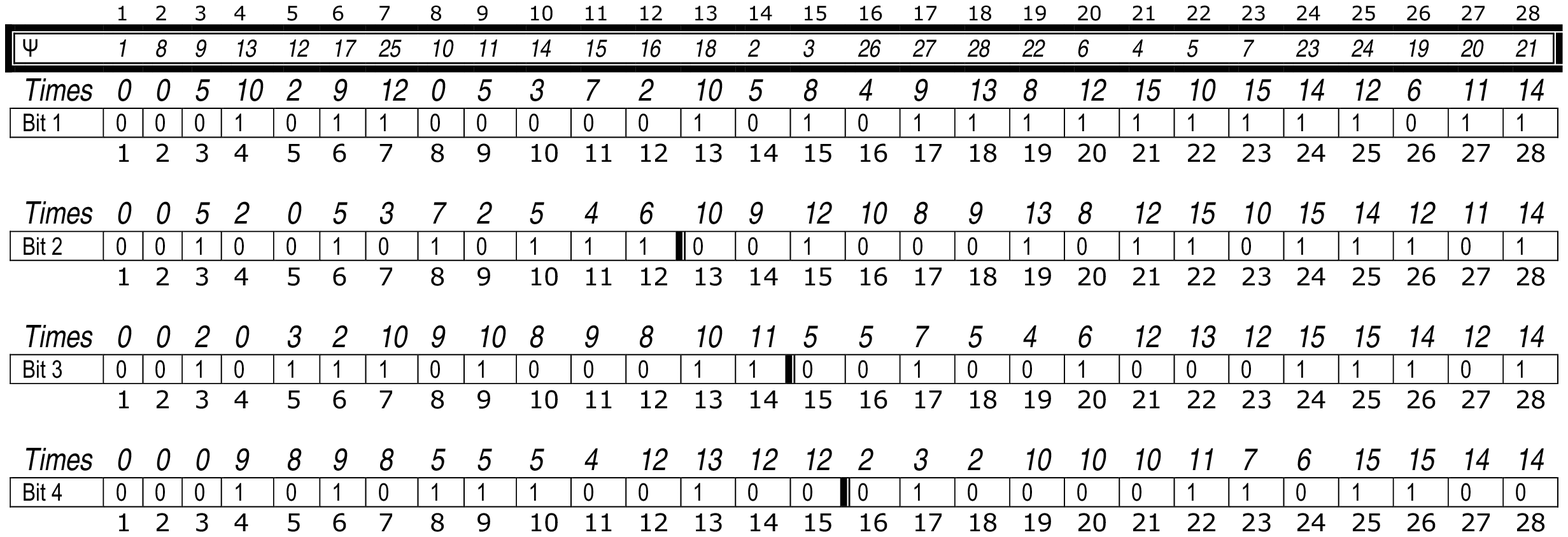}}
  \end{center}
  \vspace{-0.5cm}
  \caption{WM representation for the times associated with the trips in Figure~\ref{fig:tcsa}.}
  \label{fig:wm}
  \vspace{-0.4cm}
\end{figure}

To exploit usage patterns of a network, we need to represent
and query the time component of trips, which indicates when a moving object
reaches each node along its trip. To represent this time component,
we discretize time %instants
and assign an integer code to
each resulting time interval. The size of the time interval is a parameter
that can be adjusted to fit the required precision in each domain.
%
%Another important aspect is
%the granularity of the time period under consideration.
For example, in a public
transportation network, if we had data about five years of trips, a
possibility would be to divide that five-years period into
10-minutes intervals, or in cyclical annual periods resulting in a
vocabulary of roughly $365 \times 24 \times 60/10 = 52,560$
different codes. However,  in public transportation networks queries such
as \textit{``Number of trips using the stop X on May 10 between 9:15 and 10:00''} may be not as useful as queries such as
\textit{``Number of trips using stop X on Sundays between 9:15 and
10:00''}.
% Therefore, it is more useful
%to encode with the same codes the hours in working days on one hand and
%hours in weekend in the other.
For this reason, $\repres$ can adapt how the
time component is encoded depending on the queries that the system must answer.

In Figure~\ref{fig:tcsa}, sequence $I$ contains the time
associated with each stop in a trip, and $Icode$ a possible encoding of times.
In $\repres$ we use a similar encoding to that in
$Icode$, yet aligned to $\Psi$ rather than to $S$.%, and represented with a $\wm$.

%Each time a trip reaches a stop we assign a code as a timestamp to the stop of the trip, where timestamps are organized
%following the needs of the domain.
%In our approach,
Those entries in $Icodes$ are given a fixed-length binary code and
%codes have a fixed length and
are represented with a balanced
Wavelet Matrix ($\wm$)~\cite{CNO15}. That is, for any stop in a trip
at the position $i$ in $\Psi$, its timestamp $t_i$ can be
recovered by accessing the $\wm$ at position $i$.
Recall \cite{CNO15} that a $\wm$ is a grid of $n \times m$ bits. In  our
case $n$ is the number of entries in the $\csa$ and $m = \log  \sigma_t$ are the bits
needed to represent the different $\sigma_t$ codes for the time instants of interest.

%\marginpar{Fari recuerdame por que no te gustaba esto...}
% The codes for the timestamps must be given by taking into account the
% special sorting of codes produced by the $\wm$.
% Remember that, unlike  the WT, the $\wm$ does not sort the codes in
% lexicographic order and, therefore, we need to provide codes to timestamps in a
% way that contiguous timestamps became contiguously ordered in the
% last level of the $\wm$.

Besides the typical $access (i)$, $rank_\alpha (i)$ and
$select_\alpha (i)$, the $\wm$ provides a $count$ operation that
$\repres$ heavily relies on. $count(x1,x2,y1,y2)$ returns the number
of occurrences of symbols between y1~and~y2 in the range of
positions $[x1,x2]$ from the encoded sequence in  $O(m)$ time.
%
%In our implementation we heavily rely on the $count(x1,x2,y1,y2)$ operation from the WM,
%that returns the number of occurrences of symbols between y1~and~y2 in the range of
%positions $[x1,x2]$ from the encoded sequence.
While its implementation details can be found in~\cite{CNO15}, we include an example of how to solve
$count(20,28,10,15)$ over the sequence shown in Figure~\ref{fig:wm}.
The algorithm starts from the upper level ($Bit1$) of the $\wm$ and iterates downwards,
refining the searching range.
In $Bit1$ we are only interested in positions from $[20,28]$ that have a 1, because none
of the symbols between 10~and~15 starts with a 0.
Also, since $rank_0(Bit1,28) = 12$, in $Bit2$ we will have to search in the positions
between $12+rank_1(Bit1,20)=21$ and $12+rank_1(Bit1,28)=28$.
Now, while the second bit for 10~and~11 is 0, 
it is 1 for the symbols between 12~and~15.
Because of this, we need to perform both $rank_0$ and $rank_1$ on the limits of $[21,28]$
in $Bit2$\footnote{$rank_1(Bit2,i) = i - rank_0(Bit2,i)$, and vice versa},
and split the search in two subranges for $Bit3$: 
$[10,11]$ using $rank_0$ and $[23,28]$ using $rank_1$.
As the second subrange may only contain symbols from 12~to~15 (11xx), further refinement is not needed.
In the case of the range $[10,11]$, it could contain symbols from 8~to~11, depending on their third bits, so we need to perform $rank_1$ over its limits in $Bit3$, which leads to $[21,22]$ in $Bit4$. The number of 10~and~11 symbols is the size of this last range.

If we wanted to return the positions of the results in the original
sequence, we could do that with a simple algorithm, using
$select$ of bits  over bitmaps, that iterates upwards from the
level where each result is found  until the first level where its
position in the original sequence can be retrieved.

%If we wanted to return the positions of the results in the original sequence, we could do
%that with a simple algorithm that iterates upwards from the level where each result is found
%by using $select$ over the bitmaps.

Summarizing, $\repres$ takes the advantage of the $\wm$
to count and report the occurrences of a continuous range of values.
The starting positions in the $\csa$ belonging to the $\$$ symbols have
no time by themselves, but it is useful to answer some queries to
store the starting time instants of the corresponding trip in these
positions too.

The time intervals could be mapped to a variable-length code,
instead of a fixed length codes, where the most frequent intervals
would be represented by less bits and, therefore, requiring less
levels in a Wavelet Tree. In the future we will explore this
possibility.

%%%%%%%%%%%%%%%%%%%%%%%%%%%%%%%%%%%%%%%%%%%%%%%%%%%%%%%%%%%%%%%%%%%%%%%%%%%%%%%%%%

\section{Query processing} \label{sec:tcsaqueries}

We distinguish two types of queries to be answered by the $\repres$: spatial
and spatio-temporal queries. We briefly sketch the algorithms to process these queries.

\vspace{0.2 cm}
\noindent
\textbf{Spatial queries.} The following queries can be solved by only using the $\csa$ that represents the spatial component of trips.

%\vspace{-2mm}
\begin{itemize}[leftmargin=0mm]
\setlength{\itemindent}{0mm}
\item {\em Number of trips starting at stop $X$.}
Because $\Psi$ was cyclically built  in such a way that every $\$$ symbol is followed by the first stop of its trip, this query is solved by performing the binary search of the pattern $\$X$ over the section of $\Psi$ corresponding to \$. The size of the resulting range gives the number of trips starting at $X$.

\item {\em Number of trips ending at stop $X$.} In a similar way to the previous query, this one can be answered with a binary search for pattern $X\$$ over the section of $\Psi$ corresponding  to stop $X$.

\item {\em Number of trips starting at $X$ and ending at $Y$.}
Combining both ideas from above, this query is solved directly by searching for the $Y\$X$ pattern. % thanks to the re-arrangement of trips in $S$.

\item {\em Number of trips using stop $X$.}
Instead of performing a binary search over $\Psi$, we operate on bitmap $D$. Assuming that $X$ is
at position $p$ in the vocabulary $V$ of  $\tcsa$, its total frequency is obtained by
$occs_X \leftarrow{select_1(D,p+1) - select_1(D,p)}$. %That is, we find the range in $D$ corresponding to the stop $X$.
If $p$ is the last entry in $V$, we set $occs_X \leftarrow n+1-select_1(D,p)$.

\setlength{\itemindent}{0mm}
\item {\em Top-k most used stops.}
We provide two possible solutions for these queries: %These queries have two possible implementations:
sequential and binary-partition approaches.

\begin{itemize}
\item To return the $k$ most used stops using a sequential approach, we can
apply  $select_1$ operation sequentially for every stop from 1 to $\delta$,
returning the $k$ stops with highest frequency.
We use a min-heap that is initialized with
the first $k$ stops, and for every stop $s$ from $k+1$
to $\delta$, we compare its frequency with the frequency of the minimum stop in
the heap. In case the new one is higher, the root of the heap is replaced
and moved down to comply with the heap ordering. At the end of the process, the heap
will contain the top-k most used stops, which can be sorted with the heapsort algorithm if needed.
Note that this  approach always performs $\delta$ $select_1$
operations on $D$.

\item A binary-partition approach to solve queries about the  top-k most used stops  takes advantage of the skewed
distribution of the stops that trips visit.  Working over $D$ and $V$, $D$ is recursively split into segments of $D$ after each iteration. Each partition must, if
possible, leave the same number of different stops in each side of the partition.
The  segments created after the partitioning step are
pushed into a priority queue $Q$, storing the initial and the final positions of the segment in $D$,
and also the initial and final corresponding entries in $V$. The priority of each segment in $Q$ is
directly its size. The priority queue $Q$ is initialized with a segment covering the whole $D$ (without its initial range
of $\delta$ $\$$ symbols). When a segment extracted from the queue $Q$ represents the instance of only one stop,
that stop is returned as a result of the top-k algorithm. The algorithm stops when the first $k$
stops are found.

For example, when searching for the top-1 most used stops in the running example, $Q$ is initialized with
the segment $[8, 28]$, corresponding to stops from~1~to~10 (positions from~2~to~11 in $V$). Note
that the entries of $D$ from~1~to~7 and $V[1]$ represent the $\$$ symbol. These are not stops and
 must be skipped. Then $[8, 28]$ is split producing the segments $[8, 20]$ for stops 1~to~5
and $[21, 28]$ for stops~6~to~10. After three more iterations, we extract the segment $[14, 18]$ for
the single stop 3, concluding that the top-1 most used stop is 3 with a frequency equal to 5.
\end{itemize}
\end{itemize}

%\vspace{0.2 cm}
%\noindent
%\textbf{Temporal queries.} The following queries only use the Wavelet Matrix:
%
%\begin{itemize}
%\item {\em Number of stops used at instant $t$}. This is computed as $rank_t(n-1) -rank_t(|\mathcal{T}|)$. That is, to discard the occurrences in the $\$$ zone, we count the occurrences of instant $t$ in WM, and
%then we subtract $rank_t(|\mathcal{T}|)$, where $|\mathcal{T}|$ is the number of trips.
%
%%This can be answered simply by calculating $rank_t(n-1)$, which would return the number of occurrences of the instant $t$ in the whole WM. This number would need to be %substracted to $rank_t(|\mathcal{T}|)$, where $|\mathcal{T}|$ is the number of trips, to avoid counting occurrences in the $\$$ zone.
%
%\item {\em Number of stops used between $t_1$ and $t_2$}. The  $count$ operation from the WM can be directly applied to answer this query.
%
%\item {\em Top-k most used time intervals}. We could use a  binary partition algorithm that  performs $rank$ operations over the bitmaps of the WM. Yet, this is a slow and an unpractical approach for small time alphabets when compared with its sequential counterpart, which just requires keeping pointers to the leaves of the WM  as described in~\cite{CNO15}.
%%\footnote{Me gustaría poder hablar más sobre ambas aproximaciones, pero necesitaría un paper de revista}.
%\end{itemize}
%
%

\vspace{0.2 cm}
\noindent
\textbf{Spatio-temporal queries.} These queries combine both  the $\csa$ and $\wm$.
The idea is to restrict spatial queries to a time interval $[t_1, t_2]$.  An example of this type of query is to return  the
{\em number of trips starting at stop $X$  between $t_1$ and $t_2$}, which we solve by  relying on the  ${count}$
 %and \texttt{report} operations
operation of the $\wm$. The following are the spatio-temporal queries solved by the $\repres$:
%\marginpar{\tiny ojo, comente operacion ``report''}

%\begin{figure}[h!]
%  %\vspace{-0.4cm}
%  \begin{center}
%  {\includegraphics[width=0.95\textwidth]{figures/search.eps}}
%  \end{center}
%  \vspace{-0.3cm}
%  \caption{Trips staring at $X$ and ending at $Y$ at time interval $[t_1,t_2]$.}
%  \label{fig:search}
%  %\vspace{-0.6cm}
%\end{figure}

\begin{itemize}[leftmargin=0mm]
    \setlength{\itemindent}{0mm}
    \item {\em Number of trips starting at stop $X$ during the time interval $[t_1,t_2]$.}
    Remember that in the $\wm$ we also have timestamps associated with the area of \$-symbols in $\Psi$;
    each \$ has associated the time of the first stop of its trip and, therefore, we can use
    the $\wm$ in that area of  $\Psi$. Using the range in $\Psi$ obtained by searching the
    $\$X$ pattern, as done in a regular spatial query, a $count$ operation is performed over
    these positions in the $\wm$ searching for the limits of the interval. That is, we count the
    number of entries in the obtained range that have a timestamp in the $\wm$ inside $[t_1,t_2]$.

    \item {\em Number of trips ending at stop $X$ during the time interval $[t_1,t_2]$.}
    As before, we use a $count$ operation in the $\wm$, restricted to the range in $\Psi$ that corresponds to the pattern $X\$$ found in the spatial query.

    \item {\em Number of trips using stop $X$ during the time interval $[t_1,t_2]$.}
    As in the spatial query, the range in $\Psi$ is obtained with two $select_1$ on $D$.
    Then, a $count$ operation is done over the $\wm$ to find the occurrences inside the time interval $[t_1,t_2]$.

    \item {\em Number of trips starting at $X$ and ending at $Y$ occurring during  time interval $[t_1,t_2]$.}
    We consider two different semantics. A query with  {\em strong semantics} will obtain trips
    that start and end inside  $[t_1,t_2]$. Whereas, a query with  {\em weak semantics} will obtain trips
    whose time intervals overlap  $[t_1,t_2]$ and, therefore, they could actually start before $t_1$ or end after $t_2$.

%   When sorting  the trips, the range for $Y\$X$ in $\Psi$ matches to a continuous range of the same size in the $\$$ region ($\$XY$), preserving the same order (which is the start time of the trip). See Figure~\ref{fig:search}. Let $[i,j]$ and $[\alpha,\beta]$ be these ranges, which also  have a continuous subrange $[\alpha',\beta']$ of trips that starts inside the time interval of the query  $[t_1, t_2]$.  Note that $[\alpha',\beta']$ can be obtained as $report([\alpha,\beta],[t_1,t_2])$.
%
%   \begin{itemize}
%   \item {\em Strong semantics.} Note that subrange $[\alpha',\beta']$ (containing trips starting within $[t_1,t_2]$) has a matching subrange $[i',j'] \subseteq [i,j]$, where some of the ending times of these trips will fall inside $[t_1, t_2]$ (this allows us to check the ending time constraint). By performing a $count$ over $[i',j']$, we get the desired result. Note that this query only requires one $\Psi$ search (to find $[i,j]$), one access to $\Psi$ to obtain $\alpha$ ($\beta\leftarrow \alpha+j-i$), one $report$ to find $[\alpha',\beta']$, and one $count$ (to count the valid ending times in $[i',j']$).
%
%
%   \item {\em Weak semantics.}
%    The size of $[\alpha',\beta']$ is already a partial answer. To get the final result, we need to add occurrences of trips starting before, but ending after $t_1$. To do so, we $count$ the time instants in $[i,i')$ that fall within $[t_1, t_{max}]$. That is, ending time instants that occur after $t_1$.
%   \end{itemize}

We can binary search  $\Psi$ for the pattern $Y\$X$, hence obtaining
the corresponding continuous range of positions in the section of
$\Psi$ devoted to $Y$. We know that the range for $Y\$X$ in $\Psi$
has pointers to the section $\$$ in $\Psi$. But, note that taking
into account the considerations in the sorting of trips when
building the $\csa$, this section $\$XY$ is a continuous range of
the same size than the range  $Y\$X$, and it also preserves the same
order of the trips.

Note that, the range  $Y\$X$ of $\Psi$ has associated the final
time of each trip in the $\wm$, whereas the range $\$XY$ has associated
the timestamps of the starting time of each trip in increasing order
(due to how we sorted the trips).  Therefore, we can use these
ranges, respectively, to check time constraints related to the
ending stop ($Y$) and to the starting stop ($\$X$) of each trip.

    \begin{itemize}

    \item {\em Strong semantics.} Since time instants within the range $\$XY$ are in increasing order, we
can use the WM to obtain a continuous subrange (inside $\$XY$) of
trips starting during the interval $[t_1,t_2]$. That subrange has a
matching subrange inside of the range $Y\$X$ corresponding to the
final stop of those trips (in the same order). We can again use the
WM to count the number of those trips with valid ending times. That
is, we can perform a $count$ operation in the WM over the subrange of
$Y\$X$ corresponding to the subrange of $\$XY$ with valid starting
times.
%\marginpar{Tal y como está explicado, no nos hace falta hablar de la operación $report$ en ninguna parte del paper.}

   \item {\em Weak semantics.} In this case we need to consider all the trips in the range $\$XY$ starting within $[t_1,t_2]$, as well as the ones starting before $t_1$ but ending after $t_1$.
   %redundante
   %, and trips in the range $Y\$X$ ending after $t_2$ but starting before $t_2$.

    \end{itemize}
\end{itemize}

\section{Experimental evaluation} \label{sec:experiments}
%%%%%%%%%%%%%%%%%%%%%%%%%%%%%%%%%%%%%%%%%%%%%%%%%%%%%%%%%%%%%%%%
In this section we provide experimental results to show how $\repres$ handles a large collection of trips. We discuss both the space requirements of our representation and also show its performance at query time. 
Although due to legal issues we could not provide experiments over real trips gathered from transport companies, we managed to use real data of the Madrid's public transportation network\footnote{ Data from
the EMT corporation \url{https://www.emtmadrid.es/movilidad20/googlet.html}} (in the GTFS\footnote{GTFS is a well-known specification for representing an urban transportation network. See
\url{https://developers.google.com/transit/gtfs/reference?hl=en}} format)  to generate two datasets of synthetic trips:
\begin{itemize}
\item \textbf{Subway trips.} This combines the subway network with the Spanish commuter rail system called ``Cercanías''. In total, there are $313$ different stations organized in $23$ lines. They are open to the public from 6:00 AM to 2:00 AM, thus trips were always generated within 20 hours a day.
\item \textbf{Bus trips.} It uses  $4648$ bus stops, organized in $206$ lines. Some of these lines are from special night services, so we generated
trips using 24 hours a day.
\end{itemize}

% Esto ya no es verdad!
% In a real world scenario, some stops are more frequently used than others for both starting and
% ending a trip.
% Unfortunately, we had no data about  stops usage, so we
% managed to simulate the real behavior by following a Zipf's Law\cite{zipf1949human} to  assign
% the probability of each stop being used as either a
% starting or ending stop in a trip. In that way, the probability
% $p(k)$ for the $k^{th}$ stop to be chosen as an entry or exit stop is obtained as:
% $p(k) = \frac{1}{kH_\delta} \mbox{, where } H_\delta = \sum_{s=1}^{\delta}{\frac{1}{s}}$.

Trip generation process choses a starting stop and an ending stop, and uses the network description
to generate every stop that the trip must traverse. We generated $50$ million trips in both datasets,
whose lengths vary from $2$ to $31$ stops following a binomial distribution with a mean length of $11.81$ stops.
Based on the GTFS data, we also generated realistic timestamps along each stop, and built the $\wm$-based time index
 %described  in Section~\ref{sec:time},
in $\repres$
 discretizing these timestamps into 5-minute intervals. We distinguished four kinds of days in a week:
regular working days, Fridays/holiday eves, Saturdays, and Sundays/holidays; and two kinds of weeks for high and low season representations. In total, a time interval may belong to eight types of day.

% The plain representation of the whole dataset (regarding stop IDs as 32-bit integers and a
% separator
% integer for each trip) occupies
% 211.89MiB.

% Esto sigue siendo verdad, pero los % son diferentes para los dos datasets (a propósito).
% Our data follows a realistic use of line changes, considering that each successive change of line is less likely.
% The percentage of trips containing from $0$ to $4$ line changes is respectively
% $36.96\%, 33.05\%,  16.07\%,  9.28\%,$ and $ 4.65\%$.

%Therefore, the frequencies of each number of transfers in the generated trips can be seen in
%Table~\ref{table:transfers}.

%\begin{table}[th!]
%\vspace{-6mm}
%\begin{center}
%\caption{Frequencies of the trips by their number of transfers.}
%
%  \begin{tabular}{|l|r|r|r|r|r|}
%  \hline
%  \# of Transfers & 0 & 1 & 2 & 3 & 4 \\ \hline
%  Frequencies & 36.96\% & 33.05\% & 16.07\% & 9.28\% & 4.65\% \\ \hline
%  \end{tabular}
%
%\label{table:transfers}
%%\vspace{-10mm}
%\end{center}
%\end{table}

Below, we  show the space/time tradeoff for both datasets obtained by three settings of $\tcsa$.
We tune its  $\Psi$ sample rate parameter to values
16, 64, and 256, respectively.
All tests were run on a machine with an  Intel(R) Core(TM) i5-4440@3.1GHz
CPU, and 8GB DDR3 RAM. The operating system was Ubuntu 15.04 and the compiler gcc 4.9.2 (options -O3).

%%%%%%%%%%%%%%%%%%%%%%%%%%%%%%%%%%%%%%%%%%%%%%%%%%%%%%%%%%%%%%%%%%%%%%%%%%%%%%%%
%\subsection{Space comparison}

We compare the space usage of the stops representation in the $\tcsa$ with the space required by two baseline compressors:
{\em gzip} and {\em bzip2}. To measure the compression, we assume, as a reference, a plain representation that uses the least
amount of bits needed to represent every stop with a fixed width\footnote{9 bits/stop for subway trips, 13 bits/stop for bus trips}.
The sizes of these plain representations are 687.28 MiB for the subway trips dataset, and 992.59 MiB for the bus trips.
Note that we ignore the space needed for the representation of time intervals, as WM does not offer any compression by itself,
and needs 866.27~and~944.88~MiB for subway and bus trips, respectively.

Results regarding space usage are given in Table~\ref{table:space}.
Note that an $\icsa$  built on English text~\cite{FBNCPR12} typically
reached the compression of {\em gzip} (around 35\% in compression ratio).
As expected, the high compressibility of our sorted dataset of trips permits $\repres$ to improve those numbers with
compression ratios under 30\% in the most sparse setup,
much better than {\em gzip}, and even than {\em bzip2}.
Yet, $\tcsa$ offers also indexing features that allow us to perform efficient searches.

To provide a rough comparison with a database solution similar to {\em NETTRA} \cite{krogh2014path} we included in a table a row containing each trip ID (represented with 4 bytes), stop ID (represented with 2 bytes), and time interval (represented with 2 bytes instead of a full \texttt{datetime}). The size of the whole table was around 4505~MiB, without taking any indexes into account. Therefore, such representation would use at least more than twice the space of $\repres$ while it could not efficiently support the queries discussed in this paper.
\begin{table}[h!]
%\vspace{-6mm}
\begin{center}
\scriptsize

  \begin{tabular}{|l|r|r|r|r|r|}
  \hline
  Dataset  &
  \multicolumn{1}{c|}{$\Psi_{16}$} & \multicolumn{1}{c|}{$\Psi_{64}$} &
  \multicolumn{1}{c|}{$\Psi_{256}$} & \multicolumn{1}{c|}{gzip} & \multicolumn{1}{c|}{bzip2} \\ \hline

  Subway & \parbox[t]{1.75cm}{\centering 467.07\\(67.96\%)} & \parbox[t]{1.75cm}{\centering 249.14\\(36.25\%)} &
\parbox[t]{1.75cm}{\centering 193.10\\(28.10\%)} & \parbox[t]{1.75cm}{\centering 401.72\\(58.45\%)} &
\parbox[t]{1.75cm}{\centering 238.43\\(34.69\%)} \\ \hline

  Bus & \parbox[t]{1.75cm}{\centering 499.84\\(50.36\%)} & \parbox[t]{1.75cm}{\centering 283.12\\(28.52\%)} &
\parbox[t]{1.75cm}{\centering 227.42\\(22.91\%)} & \parbox[t]{1.75cm}{\centering 957.03\\(96.42\%)} &
\parbox[t]{1.75cm}{\centering 389.74\\(39.26\%)} \\ \hline
  \end{tabular}
%\vspace{-5mm}
\caption{Comparison on space usage for stops. Space in MiB.}
\label{table:space}
\vspace{-10mm}
\end{center}
\end{table}

%%%%%%%%%%%%%%%%%%%%%%%%%%%%%%%%%%%%%%%%%%%%%%%%%%%%%%%%%%%%%%%%%%%%%%%%%%%%%%%%

%\subsection{Time performance at queries}

To see the query performance of $\tcsa$,
%when dealing with the operations described in Section~\ref{sec:tcsaqueries},
we generated $10,000$ random queries of  each type, and
measured the average time required  to solve them.

Table~\ref{table:spatialtime} shows the results of spatial queries. Almost any query can be solved in the order of  ten $\mu$secs and the heaviest
Top-k within $m$secs per query in our experiments.
As expected, the query {\em ``ends at X''} performs slightly faster than {\em ``starts at X''}, as the region in $\Psi$ for any stop $X$ is smaller
than the region of $\$$, thus needing more time to search a pattern inside the later. It is also expected that the spatial {\em ``uses X''} performs
much faster than any other query as it does not operate over $\Psi$ and its samples, using instead the $select_1$ operator over $D$.  For the same
reasons, both spatial Top-k algorithms are also independent from the $\Psi$ sample rate parameter.
However, it is interesting to point
out that even when the binary partitioning algorithm is much faster for small values of $k$, its
sequential counterpart overcomes it %the binary search on
for large values of $k$.
This is a reasonable phenomena considering that for large values of $k$, the number of
$select_1$ operations that the binary partitioning algorithm needs to perform tends to be the
same as in the sequential algorithm, but with the additional cost of maintaining a larger and more
complex structure (a priority queue versus a binary heap).

\begin{table}[h!]
\scriptsize
\centering
\begin{tabular}{|c|c|c|c|c|c|c|c|c|}
\hline

$\tcsa$ & \parbox[t]{1.35cm}{Starts at X} & \parbox[t]{1.25cm}{Ends at X} & \parbox[t]{1.35cm}{Starts at X ends at Y} & \parbox[t]{0.90cm}{Uses X}
& \parbox[t]{1.25cm}{Sequential Top 10} & \parbox[t]{1.00cm}{Binary Top 10} & \parbox[t]{1.25cm}{Sequential Top 1000} & \parbox[t]{1.20cm}{Binary Top 1000} \\ \hline

Subway $\Psi_{16}$ & 6.03 & 4.53 & 11.24 & \multirow{3}{*}{0.3902} & \multirow{3}{*}{50.42} & \multirow{3}{*}{39.36} & \multirow{3}{*}{62.79} & \multirow{3}{*}{75.09} \\ %\hline
Subway $\Psi_{64}$ & 8.22 & 4.61 & 16.68 & & & & & \\ %\hline
Subway $\Psi_{256}$ & 18.78 & 5.69 & 38.82 & & & & & \\ \hline

Bus $\Psi_{16}$ & 7.51 & 6.27 & 9.24 & \multirow{3}{*}{0.7944} & \multirow{3}{*}{761.14} & \multirow{3}{*}{588.01} & \multirow{3}{*}{1031.84} & \multirow{3}{*}{1514.07} \\ %\hline
Bus $\Psi_{64}$ & 9.58 & 6.52 & 15.72 & & & & & \\ %\hline
Bus $\Psi_{256}$ & 22.77 & 11.35 & 41.31 & & & & & \\ \hline
\end{tabular}

\caption{Time performance for  spatial queries (in $\mu secs$/query).}
\label{table:spatialtime}
\vspace{-5mm}
\end{table}

Table~\ref{table:sttime} shows the results of spatio-temporal queries. 
Looking at the differences between spatial queries and their spatio-temporal
counterparts, it can be seen that computing a $count$ 
query  over the $\wm$ takes roughly around 3 $\mu$sec, so its time overhead is
relatively small.

\begin{table}[h!]
\scriptsize
\centering
\begin{tabular}{|c|c|c|c|c|c|}
\hline

$\tcsa$ & \parbox[t]{1.35cm}{Starts at X} & \parbox[t]{1.25cm}{Ends at X} & \parbox[t]{2.20cm}{Starts at X \\ends at Y (strong)}
& \parbox[t]{2.10cm}{Starts at X \\ends at Y (weak)} & \parbox[t]{0.90cm}{Uses X} \\ \hline

Subway $\Psi_{16}$ & 8.34 & 7.44 & 22.42 & 18.95 & \multirow{3}{*}{2.08} \\ %\hline
Subway $\Psi_{64}$ & 11.21 & 7.83 & 28.07 & 24.32 & \\ %\hline
Subway $\Psi_{256}$ & 21.68 & 8.58 & 49.98 & 46.50 & \\ \hline

Bus $\Psi_{16}$ & 10.41 & 9.50 & 12.25 & 12.12 & \multirow{3}{*}{4.90} \\ %\hline
Bus $\Psi_{64}$ & 12.95 & 10.19 & 18.84 & 18.75 & \\ %\hline
Bus $\Psi_{256}$ & 26.20 & 14.87 & 44.84 & 44.92 & \\ \hline
\end{tabular}

\caption{Time performance for spatio-temporal queries (in $\mu secs$/query).}
\label{table:sttime}
\vspace{-10mm}
\end{table}

%With respect to temporal queries, Table~\ref{table:ttime} shows the results. As you can observe all queries can be  efficiently solved.
%
%\begin{table}[h!]
%\centering
%\scriptsize
%\begin{tabular}{|c|c|c|c|}
%\hline
%
%Dataset & \parbox[t]{1.25cm}{Between $T_1$ and $T_2$\vspace{0.05cm}} & \parbox[t]{1.00cm}{Top 10} & \parbox[t]{1.15cm}{Top 1000} \\ \hline
%Subway & 1.37 & 64.90 & 167.34 \\ \hline
%Bus & 2.09 & 81.86 & 195.64 \\ \hline
%\end{tabular}
%\caption{Time performance for  temporal queries (in $\mu secs$/query).}
%\label{table:ttime}
%\end{table}

%%%%%%%%%%%%%%%%%%%%%%%%%%%%%%%%%%%%%%%%%%%%%%%%%%%%%%%%%%%%%%%%%%%%%%%%%%%%%%%%%%

%%%%%%%%%%%%%%%%%%%%%%%%%%%%%%%%%%%%%%%%%%%%%%%%%%%%%%%%%%%%%%%%%%%%%%%%%%%%%%%%
\section{Conclusions and future work} \label{sec:conclusions}
%%%%%%%%%%%%%%%%%%%%%%%%%%%%%%%%%%%%%%%%%%%%%%%%%%%%%%%%%%%%%%%%%%%%%%%%%%%%%%%%

As better tracking mechanisms will be installed, the problem of
storing and querying trips to support network analysis  will gain
interest for network management administrations and even end-user
applications. For instance, with enough data of vehicle trips from a
significant amount of drivers over the network formed by the streets
of a city, it would be possible to infer traffic rules by examining
turns that nobody takes, their usual driving speed across the
network, and other useful information.

We showed that $\repres$ is a powerful structure to represent
user trips. Using compact data structures to represent trips over a
transportation network allows us not only to keep a much larger amount
of data in main memory (compression ratio is around 30\%), but also to efficiently perform spatial and
spatio-temporal queries oriented to understand the real usage of the
network.

We have presented $\repres$ as a proof of concept development. It 
is flexible enough to allow new adaptations and functionality improvements we plan
to do as future work, such as the analysis of line changes in
switching stops (that would require storing the network  topology)
or providing compression for the time index. 
Also, future work considers providing new experiments over real data of trips.

% We assigned a range of IDs ($[1,\delta_r]$) to regular stops and a
% different one for switching stops ($[\delta_r+1,\delta]$) to provide
% a simple mechanism to know how many times a stop was used to change lines.
% For example, let us assume two lines $L_x$
% and $L_y$ whose stops are respectively $\langle ..., 25, \underline{3662}, 26,
% ... \rangle$ and $\langle ..., 116, \underline{3662}, 117,.. \rangle$, and where
% stop $3662$ is a switching stop among them that does not appear in any
% other line. To obtain the number of times stop $3662$ was used to
% change line we would  look for patterns: $\langle 25, 3662, ?x
% \rangle$, $\langle 116, 3662, ?y \rangle$, $\langle 25+1, 3662,
% ?\overline{x} \rangle$, $\langle 116+1, 3662, ?\overline{y} \rangle$
% checking that: $x \neq 26$, $y \neq 117$, $\overline{x} \neq 25$,
% and $\overline{y} \neq 116$. Obviously this is a simplification but
% a real implementation will need a proper representation of the lines
% and the network topology.

%%%%%%%%%%%%%%%%%%%%%%%%%%%%%%%%%%%%%%%%%%%%%%%%%%%%%%%%%%%%%%%%%%%%%%%%%%%%%%%%%%
%%%%%%%%%%%%%%%%%%%%%%%%%%%%%%%%%%%%%%%%%%%%%%%%%%%%%%%%%%%%%%%%%%%%%%%%%%%%%%%%%%%%%%%

\bibliographystyle{plain}
\bibliography{refs,abbrev,trajectories,secoD,BIB}

\end{document}